\documentclass[twocolumn,preprintnumbers,amsmath,amssymb]{revtex4}
\usepackage{graphicx}
\usepackage{dcolumn}
\usepackage{bm}

\newcommand{\be}{\begin{equation}}
\newcommand{\ee}{\end{equation}}
\newcommand{\bea}{\begin{eqnarray}}
\newcommand{\eea}{\end{eqnarray}}

\newcommand{\nn}{\nonumber}

\newcommand{\beq}{\begin{equation}}
\newcommand{\eeq}{\end{equation}}

\newcommand{\cL}{\mathcal{L}}

\newcommand{\cN}{\mathcal{N}}

\newcommand{\cB}{\mathcal{B}}

\newcommand{\cV}{\mathcal{V}}

\newcommand{\I}{\text{Im}}
\newcommand{\R}{\text{Re}}

\newcommand{\bbZ}{\mathbb{Z}}

\newcommand{\cref}{{\bf [check ref]}}

\begin{document}
\preprint{MPP-2014-99}
\title{Axion Inflation in F-theory
}
\author{Thomas W.~Grimm}
\affiliation{%
$^{1}$ Max-Planck-Institut f\"ur Physik, 
                F\"ohringer Ring 6, 80805 
               Munich, Germany 
}
\begin{abstract}
We study the dynamics of axion-like fields in F-theory and suggest that 
they can serve as inflatons in models of natural inflation. The axions 
arise from harmonic three-forms on the F-theory compactification space and parameterize 
a complex torus that varies over the geometric moduli space. 
In particular, this implies that the axion decay constants depend on the complex structure moduli
that can be fixed by background fluxes. 
This might allow tuning them to be super-Planckian in a controlled way 
and allow for interesting single field inflationary models. We 
argue that this requires a localization of the three-forms near 
regions of strong string coupling, analogously to the reasoning that GUT physics requires the use of 
F-theory. These models can admit a tensor to scalar ratio $r>0.1$.

\end{abstract}
\maketitle

\section{Introduction and Summary}

Inflation was proposed to solve several cosmological puzzles, including 
the homogeneity, isotropy, and flatness of the universe, as well 
as the absence of relic monopoles \cite{Guth:1980zm,Linde:1981mu,Albrecht:1982wi}. 
The simplest models 
of inflation are driven by slowly rolling scalar fields \cite{Baumann:2009ds}. 
Inflationary models also predict small inhomogeneities in the cosmic
microwave background that can be used to test the inflationary paradigm.
Recently, the \textsc{Bicep}2-experiment announced the discovery of a non-zero 
signal of primordial gravitational waves in the B-mode power spectrum \cite{Ade:2014xna}.
The measured B-mode spectrum 
 was well-fitted with a lensed $\Lambda$-CDM model with a 
tensor to scalar ratio $r=0.20^{+0.07}_{-0.05}$.
If this result survives further tests it places remarkably strong constraints 
on inflationary models. In particular, it suggests that large field 
models in which the inflaton rolls over super-Planckian distances are favored. 
To control such a scenario within an effective field theory, 
an embedding into a theory of quantum gravity such 
as string theory is desired. Such a UV completion 
allows to examine the flatness of the potential for such 
large field ranges.

There have been various suggestions how to realize an 
inflation dynamics within string theory \cite{Baumann:2009ds} and only a few can accommodate 
a large tensor to scalar ratio as has now been observed. 
One way to construct a well-controlled large field inflationary
model is to postulate that the scalar inflaton admits a classical 
shift symmetry that is preserved perturbatively. 
Such a Peccei-Quinn symmetry can naturally protect 
the two scales of inflation necessary to roll over 
a long distance in a flat potential \cite{Freese:1990rb}. 

Many candidate axions with these
properties arise in string theory as zero modes of the R-R and NS-NS 
form fields. The explicit value of the axion decay constants in string theory 
has been examined in various string compactifications. 
A systematic study appeared, for example, in \cite{Svrcek:2006yi}. Already 
earlier, it was claimed that axion decay constants are 
always sub-Planckian in string theory. More precisely, 
it was argued that in a controlled compactification one has 
to be at large volume and small string coupling, at least in some dual frame, 
which naturally leads to a suppression of the axion decay constants. 
In this work we argue that F-theory provides a 
natural set of axions that can admit large axion decay constants. 
More precisely, we claim that two crucial points about inflation in 
string theory can be successfully addressed: (1) We identify axions 
with perturbative shift-symmetries that are the lightest fields during 
inflation in a controlled setting generalizing  \cite{Grimm:2007hs}; 
(2) We argue that in F-theory field excursions can be potentially super-Planckian 
due to calculable large axion decay constants and a non-perturbatively 
induced scalar potential.

It was suggested in \cite{Grimm:2007hs} that axions arising 
from the R-R two-form in Type IIB orientifold compactifications with 
minimal four-dimensional supersymmetry might serve
as candidate inflatons. To realize such a scenario it 
was important that moduli stabilization using background fluxes 
and non-perturbative effects \cite{Grana:2005jc,Douglas:2006es,Denef:2008wq} ensures 
that these axions can be the lightest fields during inflation. Furthermore, they combine 
with the NS-NS two-form axions into complex fields $G^a$
and are therefore not directly linked to the geometric moduli 
of the compactification space. In fact, the $G^a$ parametrize a complex 
torus with metric depending on the geometric moduli of the compactification space
and the string coupling. The arguments of \cite{Banks:2003sx} state that in 
a controlled region of moduli space, i.e.~at weak string coupling and large volume, 
the metric for axions encoding the axion decay constants is 
always sub-Planckian. Therefore,  \cite{Grimm:2007hs} implemented the $N$-flation 
scenario \cite{Kim:2004rp,Dimopoulos:2005ac}, in which inflation is driven by $N$ axions using an assistance effect \cite{Liddle:1998jc,Kanti:1999ie} \footnote{See also \cite{Cicoli:2014sva} for a different attempt to realize $N$-flation in string theory.}. 

In this work we want to propose a scenario that might circumvent the 
no-go results of \cite{Banks:2003sx} by using an F-theory background. It should be stressed, however, that we do not 
expect that one can tune the axion decay constants arbitrarily large without 
ruining the inflationary potential. Instead, we suggest that the limits from the weakly 
coupled Type IIB analysis will be weakened and a wider range of possibilities for model building
is accessible in F-theory. The precise upper limits on the axion decay constants will 
depend on the form of the scalar potential, which we discuss in more detail in section \ref{sugra-embedding}
and \ref{moduli-stab}. 

\subsection{Description of the scenario: A landscape of computable 
axion decay constants} 

In order to obtain large axion decay constants, we 
propose to consider an F-theory setup. 
We first note that the axion decay constants of the R-R two-form axions are proportional to 
the string coupling. While suppressed at weak string coupling,
there can be an enhancement in backgrounds with strong 
string coupling regions. F-theory allows to geometrically describe Type IIB 
backgrounds in which the complexified string coupling $\tau = C_0 + i/g_s$ 
varies over the compact six-dimensional space \cite{Vafa:1996xn}.
 More precisely, 
one interprets $\tau$ as the complex structure modulus of an 
auxiliary two-torus, and encodes the background by an 
elliptically fibered geometry with two additional real dimensions.
Keeping $\cN=1$ supersymmetry in four-dimensions requires this 
space to be an elliptically fibered Calabi-Yau fourfold $Y_4$
with a base $B_3$. 
The space-time filling seven-branes are located at the singularities 
of the elliptic fiber, i.e.~when the torus pinches. Therefore, one finds 
that 
general F-theory backgrounds will admit regions of strong 
string coupling. Let us recall, that one cannot 
globally employ the $Sl(2,\bbZ)$-symmetry of Type IIB to exchange 
strong and weak coupling. While there exist seven-brane 
configurations that admit a genuine weak coupling limit \cite{Sen:1996vd},
this is generally not the case. In fact, when aiming 
to embed Grand Unified Theories (GUTs) into F-theory, 
an inherently non-perturbative configuration is required \cite{Donagi:2008ca,Beasley:2008dc}. 
Let us also  stress that such F-theory backgrounds automatically account for 
certain instanton corrections becoming relevant at strong coupling. 
The effective theory, at least for the moduli sector relevant in 
this work, can nevertheless be reliably calculated \cite{Grimm:2010ks}.

In the F-theory background the considered axions 
arise from harmonic three-forms on the Calabi-Yau fourfold $Y_4$
that admit one leg in the elliptic fiber. There are $N$
such forms, where $2N$ is the number of harmonic three-forms 
on $Y_4$ minus the number of harmonic three-forms on the base 
$B_3$. 
Indeed, these 
axions correspond in Type IIB to R-R and NS-NS two-form axions
as well as Wilson line moduli on seven-branes. The three-forms 
on $Y_4$ parameterize a complex torus $\mathbb{T}^{2N}_c$, defined in \eqref{complextori}, 
with metric depending 
on the complex structure moduli and K\"ahler moduli of $Y_4$.
This metric determines the axion decay constants and takes the simple form 
\beq \label{axion-summary}
   f_{ab}^2 = \frac{i}{\cV}  \int_{Y_4} J \wedge \bar \Psi_a \wedge  \Psi_b\ ,
\eeq
where $\Psi_a$ are (2,1)-forms depending on the complex structure chosen 
on $Y_4$, $J$ is the K\"ahler form, and $\cV$ is the volume of $Y_4$.
A more detailed discussion of \eqref{axion-summary} can 
be found in section \ref{InflationF-theory}.
The complex structure dependence of $\Psi_a$ allows to compute the 
axion decay constants in various regions in moduli space. 
Furthermore, the stabilization of complex structure moduli 
by fluxes has been investigated intensively \cite{Grana:2005jc,Douglas:2006es,Denef:2008wq}.
In fact, the counting of flux vacua suggests that 
there are a vast number of vacua in complex structure moduli 
space near strong curvature regions \cite{Ashok:2003gk,Denef:2004ze}. A variety of 
computable axion decay constants therefore seems attainable
even at controllably large volume. Depending on the value 
of the axion decay constants, one can implement either 
models of single axion natural inflation or $N$-flation. Clearly, 
also axion monodromy models \cite{McAllister:2008hb,Kaloper:2008fb,Flauger:2009ab,Berg:2009tg,Palti:2014kza,Marchesano:2014mla,Blumenhagen:2014gta,Hebecker:2014eua} should be 
attainable in F-theory. 

The functional dependence of the 
axion decay constants on the complex structure moduli 
can be computed for specific Calabi-Yau fourfolds using methods known from the 
computation of periods \footnote{See, for example, refs.~\cite{Mayr:1996sh,Klemm:1996ts} for the study of 
fourfold periods.}. We will not attempt to perform such a computation 
here. Instead, we will employ a local picture and motivate that 
large values can indeed arise if the axions localize in the internal 
space near certain seven-brane configurations. 

\subsection{Single axion model and GUTs on seven-branes}

The simplest models one can consider arise from 
geometries with $N=1$. In order to obtain  inflationary dynamics in such a 
setup, it is necessary to either engineer super-Planckian 
axion decay constants $f^2$, or to roll over several periods of the 
axion. The single axion now lives on a two-torus $\mathbb{T}^2_c$ with complex 
structure induced by the complex structure of the Calabi-Yau fourfold $Y_4$.
As an elliptic curve the complex structure of this $\mathbb{T}^2_c$ can be encoded by a single 
function $h$ varying holomorphically over the complex structure moduli 
space of $Y_4$. For the axion decay constant $f^2$ we 
find the relation $f^2 \propto (\I\, h)^{-1}$. While $\I\, h \approx \I\, \tau = 1/g_s \gg 1$ at weak string 
coupling, we claim that $\I\, h$ can be small when localizing the 
axion also near a strong coupling seven-brane. In order to see  
this, we will motivate that in a patch $\cB \subset B_3$ one can write 
\beq \label{sing_axion_decay}
  f^{2} \propto \frac{1}{\cV_{\rm b}} \int_{\cB} (\I\, \tau)^{-1} \, J_{\rm b}\wedge \tilde \omega^2 \ ,
\eeq
where $\tilde \omega^2$ is localizing the axion in the patch $\cB$, $J_{\rm b}$
is the K\"ahler form, and $\cV_{\rm b}$ is the volume of $B_3$.
Let us stress that the fact that the axions arise from three-forms with one 
leg in the elliptic fiber allows to localize the 
physics of these fields in the base $B_3$. 
The claim is that \eqref{sing_axion_decay} can pick up large contributions 
from strong string coupling regions. 

\begin{figure}[h!]
\centering
\includegraphics[width=5.5cm]{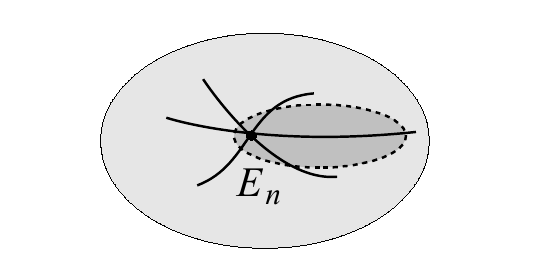} 

\begin{minipage}{8cm}\caption{Localization of the axion support (dashed line) near a 
strong coupling point.}\label{EnPoint}
            \end{minipage} 
\end{figure}

Strong coupling regions in an F-theory compactification are generic, but 
are particularly crucial for constructing F-theory GUTs. In these 
constructions the GUT group arises from a seven-brane stack on 
a four-cycle $S$ in $B_3$.
In order to obtain the necessary Yukawa structure, the geometry has to 
have points on $S$ where the local 
gauge symmetry enhances to an exceptional group $E_6,E_7$, 
or $E_8$ \cite{Donagi:2008ca,Beasley:2008dc}. One can envision 
that the axion is localized near such an exceptional point as depicted 
in Figure \ref{EnPoint}. 

Furthermore, in such a case one will also find that the complex
field $G$ containing the inflating axion $\R\, G$ will 
correct the GUT gauge coupling function as
\beq
   \alpha_{GUT}^{-1} = \text{Vol}(S) + \kappa (\I h)^{-1} (\I\, G)^2\ ,
\eeq
where $\kappa$ is a model-dependent number encoding the intersection 
number with $S$. 
This coupling has to be fixed at the GUT scale of $3 \times 10^{16}\, \text{GeV}$.
Crucially, this will also choose a frame for the $Sl(2,\bbZ)$ symmetry 
of $\mathbb{T}^2_c$ and render the notion of having large $ (\I h)^{-1}$
well-defined. In other words, as we discuss in more detail in section \ref{moduli-stab},
moduli stabilization of the the K\"ahler sector is linked to the value of $f^2$
in two ways: (1) $f^2$ directly depends on the vacuum value of the K\"ahler form $J_{\rm b}$, 
(2) the correct definition of the $\cN=1$ coordinates of the volume 
moduli is modified by $G$ and has non-trivial monodromies on $\mathbb{T}^2_c$.

\section{Inflation driven by axions}

In the following we review some basics on natural inflation \cite{Freese:1990rb,Savage:2006tr}.
We introduce the basic constructions in subsection \ref{natural}. A candidate supergravity 
embedding is described in subsection \ref{sugra-embedding}. We will argue in the next section 
that such a supergravity theory can arise in F-theory.

\subsection{Brief review of natural inflation} \label{natural}

Let us start with a set of axion-like scalars $c^a$, $a=1,\ldots, N$. 
By definition these admit a perturbatively preserved Peccei-Quinn shift symmetry
\beq \label{shift-sym}
   c^a \rightarrow  c^a + \lambda^a\ ,
\eeq
where $\lambda^a$ are constants. 
The Lagrangian for these fields takes the 
form 
\beq
  \cL = \frac{1}{2} f_{ab}^2 \partial_\mu c^a \partial^\mu c^b - V\, .
\eeq
The $f_{ab}^2$ might depend on other scalar fields of the 
theory, but are perturbatively independent of $c^a$. 
Only non-perturbative effects can induce a subleading 
$c^a$ dependence. 

To briefly discuss the phenomenological properties 
of such a model we assume that all fields determining 
$f_{ab}^2$ have been fixed to their minimum
and diagonalize $f^2_{aa} =f^2_a$. 
Canonically normalized scalars $\theta^a$ are obtained as
\beq \label{def-theta}
    \theta^a = c^a\, f_a   \ ,
\eeq
where no sum is performed.
 If the axions $c^a$ are periodic with 
period $2\pi$ the accessible field ranges are
\beq \label{field_range}
   -\pi\ <\ c^a \ \le\ \pi \ , \qquad \quad -f_a \pi \ <\ \theta^a \  \le\ f_a \pi\ .
\eeq
Since the shift symmetry \eqref{shift-sym} protects the theory 
from perturbative corrections in $c^a$, a scalar potential 
can only be induced by non-perturbative 
effects. Schematically, neglecting all cross couplings, the potential 
for the normalized axions $\theta^a$ takes the form
\beq \label{sum-pot1}
  V(\theta^a) = \Lambda_0^4 + \sum_{n^a} \Lambda^4_{n_a}\big(1- \cos \big[n_a\, \theta^a/f_a \big] \big)\ ,
\eeq
where $\Lambda_0$ is the cosmological constant 
at the vacuum $\theta^a=0$, $\{ n_a\}$ is a model dependent set of integers, and $\Lambda^4_{n_a}$
are the scales at which the Peccei-Quinn symmetries \eqref{shift-sym} are broken. One observes that the 
continuous symmetry \eqref{shift-sym} is broken by $V$ to a discrete subgroup determined 
by the set $\{ n^a \}$. 

A theory with axions $c^a$ and scalar potential \eqref{sum-pot1}
allows for models of natural inflation \cite{Freese:1990rb} or chaotic inflation for small $\theta^a$ \cite{Linde:1983gd}. 
Let us introduce the slow roll parameters for a separable potential $V$ as in \eqref{sum-pot1}.
In this case the 
slow roll parameters are given by 
\beq \label{slow_roll}
   \epsilon = \frac{M_P^2}{2} \sum_a\ \left( \frac{V_{,a}}{V}\right)^2\ ,\qquad
    \eta =    M_P^2\ \mathop{\text{min}}_{a}\left( \frac{V_{,aa}}{V}\right)\ .
\eeq
where $V_{,a}\equiv \partial_{\theta^a} V$ and $V_{,aa}= \partial^2_{\theta^a} V$, 
and $M_P =2.436 \times 10^{18}$ GeV is the reduced Planck mass. 
The slow roll conditions read $\epsilon < 1$ and $|\eta| <1$ and define a multi-dimensional 
subspace in the fields $\theta^a$ where inflation takes place. In this inflationary region of the field-space
the relevant physical observables 
can be defined as a function of the potential $V$ and its derivatives only.
For example, the tensor to scalar ratio $r$, is given by 
\beq
   r = 16 \epsilon\ .
\eeq
Therefore, it is straightforward to evaluate $r$ for the scalar potential \eqref{sum-pot1}.
As was noted in \cite{Liddle:1998jc,Kanti:1999ie,Kim:2004rp,Dimopoulos:2005ac}, even if each of the axions $\theta^a$ rolls over a 
distance smaller than  $M_P$, an assistance effect for $N$ axions 
can ensure a sufficient number of e-folds and a large $r$. 

Of particular importance in this work will be the case of one axion $\theta =c\cdot f$. 
This yields the simple and elegant model of natural or chaotic inflation.
In a general string compactification one might encounter numerous 
axions that are counted by the topological numbers of the compactification space. However, the masses of these axions can differ significantly 
during inflation, such that effectively only one field should be viewed as the inflaton.
In such a scenario, however, the axion has to roll over a distance larger than $M_P$.
In particular, calculating $r$ one finds 
\beq \label{one-r}
   r = 8 \left( \frac{M_P}{f} \frac{\text{sin}(\theta/f)}{1-\text{cos}(\theta/f)}\right)^2 \ ,
\eeq
where we have set $\Lambda_0 \approx 0$ and $f_1\equiv f,\, n^1=1$ in \eqref{sum-pot1}.
This simple formula implies that in order to have $0.1<r<0.2$ one 
has to have a super-Planckian axion decay constant $f$ \cite{Savage:2006tr}.
Furthermore, in a single field model the width of the potential must exceed 
the Lyth bound \cite{Lyth:1996im}
\beq
   \Delta \theta\  \ge \ M_P \sqrt{2\,r}\ , 
\eeq
where $\Delta \theta$ is the field excursion during inflation.

The value of $r$ is also related to the height of the potential determining 
the scale of inflation. Concretely, one finds  
\beq
V_0 = (1.94 \times 10^{16} \text{GeV})^4 \frac{r}{0.12}\ .
\eeq
This implies that for $r > 0.1$ the scale of inflation 
is at least $10^{16}\, \text{GeV}$, which is also the
GUT scale at which gauge coupling unification occurs.
Compatibility of natural inflation with the claimed 
\textsc{Bicep}2-results has been 
recently discussed in \cite{Freese:2014nla}.

\subsection{Embedding into supergravity} \label{sugra-embedding}

Before discussing axion inflation in F-theory it 
will be instructive to consider the encountered embedding 
into a purely four-dimensional supergravity setup. 
On the one hand, this will allow us to already 
comment on the properties of the appearing functions.
On the other hand, by comparing with the Type IIB 
weak coupling setup \cite{Grimm:2007hs}, we find 
that by requiring large axion decay constants, one is naturally 
led to consider F-theory setups. 

Four-dimensional $\cN=1$ supergravity 
theories require the specification of 
a K\"ahler potential $K$ and superpotential $W$ 
to encode the relevant parts of the action. 
The scalar potential is then given by
\beq \label{N=1scalarpot}
   V = e^K \big(K^{A \bar B} D_A W \overline{D_B W} - 3 |W|^2 \big)\ ,
\eeq
where $K^{A \bar B}$ is the inverse K\"ahler metric and 
$D_A$ is the K\"ahler-covariant derivative with respect to 
the complex K\"ahler coordinates.

The axions are complexified into fields $G^a =  c^a + i d^a$,
with $d^a$ real scalars in the same $\cN=1$ multiplet.
In order that the shift symmetry is preserved perturbatively,
the K\"ahler potential should only depend 
on $G^a - \bar G^a$. Indeed, in our F-theory setting with one 
K\"ahler structure modulus $T$, the K\"ahler 
potential takes the form  
\bea \label{simple-Kahlerpot}
   K &=& K_{\rm cs}  - 2 \log \cV_{\rm b} \ , \\[.1cm]
   (6 \cV_{\rm b})^{3/2} &\equiv&  T + \bar T + \tfrac14 C^{\rm cs}_{a  b}(G^a - \bar G^a) (G^b - \bar G^b) \ , \nn
\eea
where $K_{\rm cs}$ and $C^{\rm cs}_{a  b}$ generally depend on a
number of additional complex fields $z^k$, the complex structure moduli. 
In F-theory we will find that $C^{\rm cs}_{a b}$ can in fact be written 
as 
\beq \label{C=h}
  C^{\rm cs}_{a  b} =  (\I\, h^{ab})^{-1}\ ,
\eeq
where $h^{ab}(z)$ is a holomorphic function of the $z^k$. The axion 
decay constants are trivially computed from this K\"ahler potential for 
small $\R\, G^a$ to be 
\beq \label{axion-dec-super}
    f^2_{ab}  = 2 \partial_{G^a} \partial_{ \bar G^b} K = 3 \frac{(\I\, h^{ab})^{-1}}{T+\bar T} \ .
\eeq
Therefore, both the vacuum value of $\R\, T>0$ and  $\I\, h^{ab}>0$ will crucially influence the size
of the axion decay constants in this setup.

The superpotential can contain a classical piece $W_{\rm flux}(z)$ only 
depending on the $z^k$ and a non-perturbative contribution $W_{\rm np}$:
\beq \label{super1}
   W = W_{\rm flux}(z) +W_{\rm np}(z,G,T) \ .   
\eeq
One notes that the non-perturbative part depending on $G^a$ is always suppressed
by $\R\, T$ and hence should take the form 
\beq
   W_{\rm np} =  \Theta(z,G) e^{-T}\ .
\eeq
Within string theory this is not surprising, since $\R T \rightarrow \infty$
corresponds to decompactification to a higher-dimensional theory without superpotential \cite{Witten:1996hc,Ganor:1996pe,Grimm:2007xm,Grimm:2011dj}.
We will discuss the form of $ \Theta(z,G)$ in section \ref{moduli-stab}, and only make 
some general remarks in the following. Since 
$\I\, G^a$ has a Peccei-Quinn shift symmetry, it should only depend on $e^{i n_a G^a}$,
i.e.~arise from non-perturbative effects. 
This implies that after fixing $z^k$ and $T$ to their vacuum values,
one encounters an effective  
non-perturbative superpotential  
\beq \label{effesuper}
   W_{\rm eff}(G^a) = \sum_{n_a}  \Lambda^2_{n^a} e^{i n_a G^a} \ ,
\eeq
where one sums as over a set of integer vectors $\{ n^a\}$. 
The effective superpotential \eqref{effesuper} allows to induce 
the scalar potential \eqref{sum-pot1} for the axions $c^a = \R\, G^a$.
In order to realize a simple natural inflation or $N$-flation model 
it is crucial that the $\Lambda^2_{n^a} $ and the exponential
$e^{- n_a \I\, G^a}$ sufficiently suppress 
higher harmonics for the $c^a$. 

Before turning to the construction 
of the described supergravity theory in F-theory, 
it is worth making the following simple observation.
As we will see in the next section, the effective theory 
is only valid for sufficiently large $\R\, T$, but it can equally well 
be applied for all values of $\I\, h^{ab} >0$. An interesting regime 
to consider is therefore  
\beq \label{small-Imtau}
  \I\, h^{ab} \ll 1\ ,
\eeq
since it allows the axion decay constants \eqref{axion-dec-super} to be large, in 
fact, potentially super-Planckian. If this regime can be reliably 
approached in a string theory construction, a single axion model 
can implement inflation with a large $r$ given in \eqref{one-r}.
However, deriving the above data in weakly coupled Type IIB 
string theory, one finds $ \I\, h^{ab} \propto 1/g_s$, where $g_s$
is the string coupling constant. Therefore, one has to 
leave the weak coupling regime to allow for \eqref{small-Imtau}. Here
F-theory comes into the game and we suggest that 
$f_{ab}^2 >  M^2_P$ implies that one has to use a full-fledged F-theory setting.
It is interesting to remark that this is analogous 
to the fact that Grand Unified Theories cannot 
be studied in weakly coupled Type IIB string theory.

\section{Inflation from F-theory axions} \label{InflationF-theory}

Let us consider F-theory compactified on an elliptically fibered 
Calabi-Yau fourfold $Y_4$ with base threefold $B_3$.
This theory corresponds to Type IIB string theory on $B_3$ with 
seven-branes located at the singularities of the elliptic fibration. 
The four-dimensional effective theory is minimally supersymmetric,
such that its specifying data include a K\"ahler potential and superpotential.
For the analysis of the moduli action it will be sufficient to apply the 
results of \cite{Grimm:2010ks}.

\subsection{F-theory axions and their decay constants}

Crucial for our F-theory models is the fact that the effective four-dimensional 
theory arising from such a reduction admits 
\beq
   N = h^{2,1}(Y_4)-h^{2,1}(B_3)
\eeq
complex scalars $G^a$ \cite{Denef:2008wq,Grimm:2010ks}. Here $h^{p,q}(Y_4)$, $h^{p,q}(B_3)$  
are the Hodge numbers of the manifolds $Y_4$, $B_3$. To 
simplify our analysis we consider base manifolds $B_3$
with $h^{2,1}(B_3)=0$ in the following. Furthermore, recall 
that $h^{3,0}(Y_4) = 0$ for every Calabi-Yau fourfold that yields 
an $\cN=1$ effective theory in four space-time dimensions. 
To define the $G^a$ we use the dual M-theory picture, and expand the 
M-theory three-form $C_3$ as 
\beq  \label{C3exp}
    C_3 = i G^a \bar \Psi_a -i \bar G^a  \Psi_a\ ,
\eeq
where the $\Psi_a,\ a =1,\ldots, N$ form a basis of $H^{2,1}(Y_4)$.
Since the $\Psi_a$ have one leg in the elliptic fiber, performing the 
M-theory to F-theory limit shows that the fields $G^a$ correspond to 
modes of the R-R and NS-NS two-forms and Wilson line moduli 
of seven-branes \cite{Grimm:2010ks}.

{}From \eqref{C3exp} one realizes that the $G^a$ 
are coordinates on a complex $N$-dimensional  torus 
\beq \label{complextori}
     \mathbb{T}_c^{2N} = H^{2,1}(Y_4) / H^3(Y_4,\bbZ)\ ,
\eeq
which varies over the geometric moduli space of the manifold $Y_4$.
The complex structure on $\mathbb{T}_c^{2N}$
will be induced by the complex structure on 
$Y_4$ and hence will vary over the space of 
complex structure deformations of $Y_4$. 
The metric $G_{a  b}$ on $\mathbb{T}_c^{2N}$ 
takes the form 
\beq \label{axion-metric}
  G_{a b} =  \frac{1}{2\cV} \int_{Y_4}  \bar \Psi_a  \wedge * \Psi_b = \frac{i}{2\cV} \int_{Y_4} J\wedge \bar \Psi_a \wedge   \Psi_b  \ .
\eeq
where $\cV=\int_{Y_4} J^4$ is the volume and $J$ is the K\"ahler form of $Y_4$.
This metric encodes the four-dimensional kinetic terms of $G^a$ 
as $\cL_{kin} = G_{ab} \partial_\mu G^a \partial^\mu \bar G^b$.
One notes that $G_{a  b}$ depends on the complex 
structure deformations through $\Psi_a$ and 
the K\"ahler structure deformations through the 
appearance of $J$ and $\cV$.
Importantly, one expects that one can thus follow 
 $\mathbb{T}_c^{2N}$ into the interior of the 
complex structure moduli space of $Y_4$. 

The metric \eqref{axion-metric} can be derived from 
a K\"ahler potential \cite{Grimm:2010ks}, which for the simple case of having $h^{1,1}(B_3)=1$ is of the form \eqref{simple-Kahlerpot}. 
In this case the K\"ahler form $J$ on the base $B_3$ is proportional 
to the single harmonic form $\omega_{\rm b}$. 
Comparing the metric \eqref{axion-metric}
with \eqref{axion-dec-super} and \eqref{simple-Kahlerpot} one finds 
\beq  \label{CPsi}
   C^{\rm cs}_{ab} =  2 i \int_{Y_4} \omega_{\rm b} \wedge \bar \Psi_a \wedge   \Psi_b\ .
\eeq
The complex structure dependence of $C^{\rm cs}_{ab}$ therefore
arises from the fact that the notion of $\Psi_a$ being a $(2,1)$-form 
depends on the choice of complex structure on $Y_4$.

The form \eqref{C=h} of $C^{\rm cs}_{ab}$ can now be 
inferred as follows. Let us introduce a real symplectic 
basis $(\alpha_a,\beta^b)$ on $H^{3}(Y_4,\bbZ)$ satisfying 
\bea \label{alpha_beta_basis}
     &&\int_{Y_4} \omega_{\rm b} \wedge \beta^b \wedge \alpha_a  = \delta_a^b\ ,\\
     &&\int_{Y_4} \omega_{\rm b} \wedge \alpha_a \wedge  \alpha_b =     \int_{Y_4} \omega_{\rm b} \wedge \beta^a \wedge  \beta^b = 0\ . \nn
\eea
The $(2,1)$-forms $\Psi_a$ can be expanded in this basis 
as 
\beq \label{psi_expand}
    \Psi_a = \tfrac{1}{2} (\I h^{ab})^{-1} (\beta^b - h^{bc} \alpha_c)  \ . 
\eeq
where 
$h^{ab}$ is a symmetric matrix that 
depends holomorphically on the $h^{3,1}(Y_4)$ complex structure deformations $z^k$.
This can be justified by the fact that $H^{2,1}(Y_4)$ can be 
chosen to vary holomorphically over the space of complex structure deformations.
Indeed, setting 
\beq \label{def-psi}
   \psi^a = 2\, \I h^{ab} \Psi_b=\beta^b - h^{bc} \alpha_c\ , 
\eeq
one finds $\partial_{\bar z^k} \psi^a = 0$, for a holomorphic $h^{bc}(z)$. 
Inserting the expansion \eqref{psi_expand} into \eqref{CPsi}
one evaluates by using \eqref{alpha_beta_basis} that $C^{\rm cs}_{ab} = (\I h^{ab})^{-1}$,
just as claimed in \eqref{C=h}.

In order to implement axion inflation one crucially 
has to show that the classical K\"ahler potential is 
independent of the axions $\R\, G^a = c^a$. 
This is true in the above setup, since $\I\, \Psi_a$ is independent of the complex 
structure moduli. Expanding the M-theory 
three-form $C_3 = \tilde b_a \beta^a - \tilde c^a \alpha_a $ in the basis \eqref{alpha_beta_basis}
one finds by comparison with \eqref{C3exp} and using \eqref{psi_expand} 
that the complex coordinates are defined as  
\beq
     G^a = \tilde c^a - h^{a b} \tilde b_a \ ,
\eeq
and the shift symmetry of $C_3$ translates to a shift-symmetry of $c^a = \tilde c^a - \R\, h^{ab}\tilde b_b$.
This justifies the simple form of the K\"ahler potential \eqref{simple-Kahlerpot}.
A detailed derivation can be found in \cite{Grimm:2010ks}. 

Let us note that using the axions $c^a$ was 
first suggested in the weakly coupled Type IIB $N$-flation scenario of \cite{Grimm:2007hs},
and later used in axion monodromy inflation \cite{McAllister:2008hb,Flauger:2009ab,Berg:2009tg,Palti:2014kza,Marchesano:2014mla}.
At weak coupling the axions $c^a$ can be the zero-modes 
of the R-R two-forms $C_2$, and the coordinates $G^a$ are 
given by $G^a = \tilde c^a - \tau \tilde b^a$.
In this case one finds 
\beq 
   h^{a b} = \tau \delta^{ab}\ , 
\eeq
where $\tau = C_0 + i e^{-\phi}$ is a four-dimensional complex scalar 
field comprising the R-R zero-form $C_0$ and the dilaton $\phi$. 
As stressed before, at weak coupling $\I\, \tau \gg 1$ and 
the axion decay constants are naturally sub-Planckian. Inflation 
then requires the use of an assistance effect \cite{Liddle:1998jc,Kanti:1999ie}, as 
suggested for the $N$-flation scenario of \cite{Kim:2004rp,Dimopoulos:2005ac}.

Let us close this section by noting that $h^{ab}(z)$ can be computed
explicitly for a given Calabi-Yau fourfold 
example. The required techniques are similar 
to the ones for computing the complex structure dependent 
periods of the $(4,0)$-form $\Omega$ on $Y_4$.
In fact, by considering the variation of Hodge-structures, 
a basis $\psi_a$ of $H^{2,1}(Y_4)$ varying holomorphically 
over the complex structure moduli space is 
expected to satisfy a second order homogeneous differential 
equation in the complex structure moduli $z^k$. This 
should allow to derive the moduli dependence of $\psi_a$ 
explicitly. While it would be interesting to do that, we will 
take a more local route in the following to infer properties 
about $h^{ab}$. This will also hint to the fact that 
$Y_4$ should admit certain singularities that lead to four-dimensional 
gauge groups.

\subsection{Decay constant in a one axion model}

Let us next discuss the simplest possible model 
and assume that the geometry satisfies 
\beq
   h^{2,1}(Y_4) = 1\ ,  \qquad h^{1,1}(B_3) = 1\ ,
\eeq
with $h^{2,1}(B_3) =0$ as above. In this case one 
finds a single axion $c \equiv c^1$ as the real 
part of a single complex field $G$.

The complex field $G$ parameterizes a complex one-dimensional  
torus  $\mathbb{T}^2_c$ as defined  \eqref{complextori}. Recall the basic 
fact that such a torus can be mapped to 
an elliptic curve:
\beq \label{ellcurve}
   \mathbb{T}^2_c\ :\qquad \tilde y^2 = 4 \tilde x^3 - g_2(z) \tilde x - g_3(z)\ , 
\eeq
by using the Weierstrass $\wp$-function $(\tilde x,\tilde y) = (\wp,\wp')$. The coefficient functions $g_2,g_3$
depend on the complex structure induced on $\mathbb{T}^2_c$, and 
hence on the complex structure moduli $z^k$ of $Y_4$. 
To relate the $g_2(z),g_3(z)$ to the holomorphic function $h(z)$ introduced 
in \eqref{psi_expand}, we note that $\psi$ defined in \eqref{def-psi} is given by 
$\psi = \beta - h \alpha$, where $(\alpha,\beta)$ span the lattice $H^{3}(Y_4,\bbZ)$. 
Therefore, again applying a standard fact about elliptic curves, one finds the relation 
\beq \label{jfunction}
    j(h) = 1728 \frac{g_2^3}{g_2^3 - 27 g_3^2}\ ,
\eeq
where $j$ is the $j$-invariant of the elliptic curve.
The axion decay constant for the single axion is given by 
\beq
     f^2 = 3 \frac{(\I\, h)^{-1}}{T+\bar T}\ .
\eeq
The key question is whether the 
function $\I\, h$ of the complex structure moduli 
of $Y_4$ can be small such that $f$ becomes large.

In order to proceed we aim to relate the 
elliptic curve \eqref{ellcurve} to the elliptic fiber of the Calabi-Yau 
manifold $Y_4$. Let us first recall some basic facts about 
elliptic fibrations. The equation describing $Y_4$
can be brought into the Weierstrass form 
\beq
   Y_4\ :\qquad  y^2 = 4 x^3 - f(u,z)  x - g(u,z)\ . 
\eeq
In contrast to  \eqref{ellcurve}, however, the coefficient functions depend
both on the coordinates $u$ on $B_3$ and on the complex
structure moduli $z^k$. The complex structure $\tau$ of the 
elliptic fiber is given by $j(\tau)$ expressed as a function 
of $f,g$ as in \eqref{jfunction} with $g_2\rightarrow f$ and $g_2 \rightarrow g$.
Clearly, also $\tau$ depends both on the coordinates $u$ on $B_3$
and the complex structure moduli $z^k$.

In order to relate $\tau$ and $h$ we employ a local 
picture. The $(2,1)$-form $\Psi$ can be locally written as 
\beq
   \Psi = \tfrac12 (\I\, \tau)^{-1}\,  \tilde \omega \wedge (dx - \tau dy)
\eeq
where $\tilde \omega$ is a two-form supported on a patch 
$\cB$ in the base $B_3$. 
Both $\tilde \omega$ and $\tau$ depend on the location 
on $B_3$ and the integral \eqref{axion-metric} reduces on $\cB$ as
\beq \label{axion-text}
   f^2 = \frac{1}{\cV_b} \int_{\cB} (\I\, \tau)^{-1} J_{\rm b}\wedge \tilde \omega^2 \ ,
\eeq
where we have integrated over the torus fiber with normalized volume form $dx \wedge dy$
and localized to $\cB$. 
In \eqref{axion-text} we further use that the base K\"ahler form $J_{\rm b}$ and volume $\cV_{\rm b}$  arise 
when taking the M-theory to F-theory limit $\cV^{-1} J|_{B_3}  \rightarrow \cV_{\rm b}^{-1} J_{\rm b} $ \cite{Grimm:2010ks}.
One concludes that the axion decay constants 
can indeed gain large contributions near strong coupling regions 
in the base $\cB$. Interestingly, this is a local configuration and 
we believe that an explicit computation for an appropriate  
seven-brane configuration will confirm this result. 

Let us observe that \eqref{axion-text} at first seems ill-defined, since $\tau$ 
is only defined up to an overall $Sl(2,\bbZ)$ monodromy of the elliptic fiber. In 
fact, such a monodromy can exchange a D7-brane into a general $(p,q)$-seven-brane. 
This symmetry translates into the $Sl(2,\bbZ)$ symmetry of $\mathbb{T}^2_c$. 
In the vacuum, however, a frame for this symmetry is chosen when considering a
concrete moduli stabilization scenario. Indeed, the definitions of the $h^{1,1}(B_3)$ complex 
fields $T_\alpha$ containing the cycle volumes of $Y_4$ are 
non-trivially shifted by $G^a$ \cite{Grimm:2010ks}. For $h^{1,1}(B_3)=1$ one finds 
\beq
   T = \cV^{2/3}_{\rm b} + i \rho -\tfrac14 C^{\rm cs}_{ab} G^{a} (G^b-\bar G^b)\ ,
\eeq
where $\rho$ is the R-R four-form axion. 
Hence, a general $Sp(2N,\bbZ)$ symmetry transformation of $\mathbb{T}^{2N}_c$ will non-trivially shift the 
K\"ahler coordinates $T_\alpha$. Recall that also the GUT gauge coupling 
is given by the combination $n^\alpha_{\rm GUT}\, \R T_\alpha$, where the constant 
vector $n^\alpha_{\rm GUT}$ determines the location of the GUT-brane in $B_3$. Hence, fixing the 
gauge coupling on the observable brane will equally choose a frame for the symmetry of $\mathbb{T}^{2N}_c$.
In the final subsection we will collect further comments on moduli stabilization 
and the generation of a potential for $G^a$.

\subsection{Remarks on moduli stabilization} \label{moduli-stab}

In the final part of this letter we comment on moduli 
stabilization and the generation of an axion potential. 
To begin with, note that the complex structure moduli of $Y_4$ can be stabilized 
using background four-form fluxes $G_4$ on $Y_4$. 
In the  four-dimensional effective theory the scalar 
potential arises from a superpotential \cite{Gukov:1999ya}
\beq \label{GVWsuper}
     W(z) = \int_{Y_4} \Omega(z) \wedge G_4
\eeq
where $\Omega$ is the $(4,0)$-form on $Y_4$ 
depending on the complex structure 
deformations $z^k$. Moduli stabilization 
with fluxes was studied intensively \cite{Grana:2005jc,Douglas:2006es}.
In particular, it was argued in \cite{Ashok:2003gk,Denef:2004ze}
that a large fraction of the vacua derived from 
\eqref{GVWsuper} are in the interior of the
complex structure moduli space.

Clearly, in addition to stabilizing the complex structure 
moduli $z^k$ using \eqref{GVWsuper} one also needs 
to find a potential for the K\"ahler structure moduli $T_\alpha$. 
Following the suggestion of \cite{Kachru:2003aw}, one can stabilize 
these fields using a non-perturbatively generated 
superpotential 
\beq \label{W(T)}
   W(T) = \sum_{n_\alpha} \Theta_{n_\alpha}\, e^{-n^\alpha T_\alpha }\ ,
\eeq
where the sum runs over a model dependent set of 
integers $n_\alpha$ classifying allowed non-perturbative configurations
in F-theory \cite{Witten:1996bn}.

The coefficient functions $\Theta_{n_\alpha}$ can, in general, 
holomorphically depend on all other complex scalar 
fields of the theory. In particular, they can be a non-trivial 
functions of the complex structure deformations $z^k$
and the fields $G^a$. The precise definition of $T_\alpha$, which 
arises from the world-volume action of a brane instanton, suggest that 
the $\Theta_{n_\alpha}(z,G)$ are theta-functions on the torus $\mathbb{T}^{2N}_c$
\cite{Witten:1996hc,Ganor:1996pe,Grimm:2007xm}.
For example, for a single K\"ahler modulus $\Theta(z,G)$ takes the form 
\beq \label{Theta-def}
    \Theta(z,G) = f(z) \sum_{n_a \in \Gamma} e^{\frac{1}{2} i h^{ab} n_a n_b + i n_a G^a}\ , 
\eeq
where $\Gamma$ is some model dependent integer lattice. For instance, 
$\Gamma$ can be the lattice of supersymmetric fluxes supported on the brane 
instanton, as discussed e.g.~in \cite{Grimm:2011dj,Kerstan:2012cy,Cvetic:2012ts}.

In order to obtain the potential for the fields $G^a$ we 
make the following observations. There are two contributions 
to the scalar potential for $\I\, G^a$. Firstly, there is a term 
arising from expanding the prefactor $e^K$ in the $\cN=1$
scalar potential \eqref{N=1scalarpot} as
\beq
   e^K = e^{\langle K\rangle} \Big( 1 + 2 f^2_{ab}\, \I\, G^a \ \I\, G^b + ...\Big)\ ,
\eeq
where we have inserted the K\"ahler potential \eqref{simple-Kahlerpot} and $f^2_{ab}$ given in \eqref{axion-dec-super}, 
and expanded for small $\I\, G^a$. One observes that the arising term can give a significant contribution
to the mass of $\I\, G^a$
when $f^2_{ab}$ is large. Secondly, the dependence of the superpotential \eqref{W(T)} with \eqref{Theta-def} on the 
fields $\I\, G^a$ induces additional contributions to the scalar potential. 

In contrast, the K\"ahler potential is independent of $c^a=\R\, G^a$. Therefore, a periodic potential of 
the form \eqref{sum-pot1} is induced by the  \eqref{W(T)} with \eqref{Theta-def}. 
In order to build models of natural inflation or $N$-flation it will be crucial to ensure that 
higher harmonics for the $c^a$ are suppressed. Investigating simple toy examples, one realizes
that this is challenging for large $f^2_{ab} \propto (\I\, h^{ab})^{-1}$ in \eqref{Theta-def}, and 
that there will be a model-dependent upper bound on $f^2_{ab}$ with suppressed higher harmonics.

\section{Conclusions}

In this letter we have studied models of natural inflation realized in F-theory. The inflatons
were chosen to be axions arising from harmonic three-forms with one leg in the elliptic fiber
of the compactification Calabi-Yau fourfold. In the 
four-dimensional $\cN=1$ effective theory the inflating axions are the real parts of
complex fields $G^a$. These are singled out, since 
only $\I\, G^a$ appear in the K\"ahler potential of the theory.  
 The axions correspond to 
R-R or NS-NS two-form axions, and seven-brane Wilson line axions in Type IIB string theory.    
The three-forms span a complex $N$-dimensional 
torus $\mathbb{T}^{2N}_c$ with complex structure and metric depending on the 
geometric moduli of $Y_4$.
In contrast to weakly coupled Type IIB compactifications,
the F-theory axion decay constants are non-trivial functions over the complex structure 
moduli space of $Y_4$.  It is conceivable that a systematic fixing of 
complex structure moduli using the well-known flux superpotential will
allow for a rich value distribution of axion decay constants. 
It would be desirable to compute the complex structure dependence 
explicitly for specific Calabi-Yau fourfold examples. While this 
is expected to be possible for Calabi-Yau fourfolds with few 
complex structure moduli, it will be challenging to perform such 
computations for fourfolds with GUT singularities. For 
such situations a local approach would be desirable at first.

The main claim of this paper is that F-theory provides the opportunity to 
access possibly super-Planckian axion decay constants 
in a controlled way including $g_s$-dependent instanton corrections. 
We argued that large contributions to the axion decay constants 
arise when the axions are localized in strong coupling regions 
on the base space $B_3$. Strong coupling effects are necessary to 
realize GUTs in F-theory, but could also occur near a hidden seven-brane.
To establish an invariant notion of strong coupling it was also crucial to fix 
the symmetries of $\mathbb{T}^{2N}_c$ that we argued to be linked to 
the overall $Sl(2,\bbZ)$ symmetry of the F-theory setup. 
The latter symmetry can be used to locally
transform strong into weak string coupling. If the axions 
are localized near complicated seven-brane configurations, however, 
this symmetry might only permute which brane-regions   
contribute dominantly to $f^2$. Furthermore, a frame for the symmetries of $\mathbb{T}^{2N}_c$
is chosen once moduli stabilization is implemented. In 
particular, expanding the non-perturbative superpotential and keeping only 
the leading terms in the K\"ahler structure moduli $T_\alpha$ and the 
$G^a$ will break the symmetries. 

Clearly, it remains to be an important open  task to find concrete examples with 
large axion decay constants $f_{ab}^2$ and a sufficiently flat scalar potential. 
One challenging part in this endeavour will be to idenfy models with
scalar potentials arising from instanton effects that 
sufficiently suppress higher harmonics for the inflations $c^a$.
The success will depend on the flux-lattice $\Gamma$ appearing in \eqref{Theta-def}
and the vacuum expectation values for the stabilized moduli including $\I\, G^a$. 

Let us end with a final comment concerning reheating 
in these setups. At the end of natural inflation the axions oscillate around 
their minimum and decay into various coupled modes \cite{Freese:1990rb,Savage:2006tr,Freese:2014nla}. In particular, 
a coupling $m_a c^a  \text{Tr} (F \wedge F)$ can 
yield decays into the GUT or a hidden sector gauge fields. 
In our setting $m_a$ is a seven-brane flux. An appropriate choice of 
$m_a$ therefore allows to control the size of the decays in this channel,
similar to the discussion of \cite{Blumenhagen:2014gta}. It would be interesting to 
study reheating in this F-theory setting in detail.

\vspace*{.5cm}

{\noindent \bf Acknowledgments:}
I would like to thank Ralph Blumenhagen, Federico Bonetti, I\~naki Garc\'ia-Etxebarria, Jan Keitel, Eran Palti, Tom Pugh, and Edward Witten for interesting discussions 
and comments on the draft. This research was funded by a research grant of the Max Planck Society.

%
%

\end{document}